\documentclass[aps,prl,twocolumn,preprintnumbers]{revtex4}
\usepackage{graphicx}
\begin{document}
\newcommand{\BM}[1]{{\mbox{\boldmath{$#1$}}}}
\newcommand{\vev}[1]{\langle{#1}\rangle}

%\hfill\vbox{ \hbox{hep-ph/0403070} \hbox{}}

%\preprint{\sf Version 1 (\today)}
\title{Natural Kobayashi-Maskawa Model of CP Violation and Flavor Physics }
\author{Darwin Chang}
%\altaffiliation[]{Authors listed in alphabetic order.}
\affiliation{Physics Department and NCTS, National Tsing-Hua University,
Hsinchu 300, Taiwan}

\author{Wai-Yee Keung}
\affiliation{Department of physics, University of Illinois at
Chicago, Illinois 60607-7059, USA}

\begin{abstract}
We explore the possible tie between the naturalness of having a
very small strong CP $\theta$ parameter in the Kobayashi-Maskawa
Model and the flavor symmetry.  We provide many examples in which
the flavor symmetry group at high energy can naturally give rise
to Kobayashi-Maskawa Model at low energy with a naturally small
$\theta$.
\end{abstract}

\author{(Contribution delivered by DC to ICFP2003, KIAS, Seoul, Korea, 
to appear in Korean Journal of Physics)}
\noaffiliation

% \pacs{Valid PACS appear here} \maketitle %%
\maketitle

\section*{Introduction}

Besides some recent anomalies at the two B factories, the Standard
Kobayashi-Maskawa (KM) Model of CP Violation has been amazingly
successful.  Flavor physics is at the center of the
Kobayashi-Maskawa(KM) model, the standard model of CP violation.
The origin of flavor mixing is a mismatch between the mass
eigenstates and the weak gauge eigenstates.  Therefore, within the
Standard model, the origin of flavor is the Yukawa couplings which
give rise to all fermion masses. In KM model, the CP violation is
mediated by the charged current processes that interfere with each
other. With three generations, the charge current mixing matrix,
the CKM matrix, has exactly one physical phase. Indeed it was
first observed by Kobayashi and Maskawa(KM)\cite{Kobayashi:fv}
that the two generation Standard Model cannot support any CP
violating phase. (Actually Ref.\cite{Kobayashi:fv} proposed many
different additions to the Standard Model in order to incorporate
CP violation, including the KM model and Higgs models).  The fact
that all three generations have to be involved to create a CP
violating phenomena, makes KM model extremely subtle and beautiful
model for CP violation.  It also makes CP violation tightly
connected with flavor physics. This partly explains why the CP
violation is small in the kaon system. It is because, at the
energy scale of kaon, the heavy third generation almost decouples
and, as a result, the effective CP violation is more
superweak-like despite the fact the phase in the CKM matrix may
not be small at all in some convention (In particular, in
Wolfenstein\cite{Wolfenstein:1983yz} or in
Chau-Keung\cite{Chau:fp} conventions).  This difference in the
quark masses is of course related to the hierarchy in the
eigenvalues of the Yukawa coupling matrices which are bare
parameters in Standard Model.

The situation of course changes in the $B$ meson system.
The higher masses of $B$ (and $B_s$) mesons open up many
decay channels of different flavor property and make it a great
laboratory to study flavor physics and the associated CP violating effect.
As a result, there has been a large amount of literature studying the
effect of CP violation in different $B$ decay
channels and we shall not dwell into this.  It suffices to say that KM
model predicts special pattern of CP violating effects in various $B$ (and
$B_s$) decay channels which can be contrasted with the patterns arisen
from other CP violating mechanisms in the near future.
It is also important to note that any theory that attempts to explain
the origin of fermion mass hierarchy will automatically alter the picture of
CP violation puts forth in KM model for better or for worse.

There are two often quoted weak points in the KM mechanism.
First, it has been shown that KM model alone is insufficient to produce large
enough baryon asymmetry in the cosmic evolution.
The main difficulties are two folds.
First of all, given the current limit of the Higgs mass, the
electroweak phase transition tends to be not strongly first order
enough to provide the off-equilibrium condition necessary to
generate baryon asymmetry.
Secondly, the CP violating source in CKM matrix tends to be too
small to provide large effects.
There are various attempts to extend KM model to overcome the above
difficulties and to generate the baryon asymmetry at electroweak scale.
Two leading
proposals are (1) the minimal supersymmetric extension of Standard Model
(MSSM), or
(2) two Higgs doublet extension.  Of course, the alternative is to give up
electroweak
baryogenesis and produce baryon or lepton asymmetry at higher energy scale.

The second weakness of KM mechanism is that it does not explain why
strong CP $\theta$ is so small.  There are two levels to this problem, the
tree and the loop levels.  Since the CP violating phase in KM model is a
part of the
Yukawa couplings, which are dimension four couplings just like the tree
level $\theta$ parameter, the $\theta$ should naturally be one of the bare
parameters of the model with a natural value of order one.
The phase convention independent, physical $\theta$, which is a linear
combination of three level $\theta_0$ and extra contribution from the
quark mass matrices, $\theta =
\theta_0 - Arg(det{M^u M^d})$, should also be naturally of order one.
This is the tree level strong CP problem.
The second issue is, even if the $\theta$ is tuned to zero at tree level,
radiative corrections still make it too large or divergent.
However, it should be noted that, by being tightly connected with the
flavor physics as described above,
the KM mechanism has already embedded in itself a natural mechanism to
suppress the loop correction to $\theta$.  It has been shown that if
one heuristically set the physical $\theta$ to zero at the tree level,
nonzero loop correction will not occur till three loop level
(with two weak and one strong loops)\cite{Shabalin:zz},
and logarithmic divergent correction to
$\theta$ can appear only at the 14th order of the electroweak coupling $g_2$
(or, at the 7-loop level)\cite{Ellis:1978hq,Khriplovich:1993pf}.
Even if one puts in the Planck scale as the estimate of the cut-off, the
divergent correction produces only a minute value for $\theta$ just like
the 3-loop finite corrections.  That is, the special $\theta =0$ point,
while not natural, is actually quite robust under radiative corrections.
This nice property is a
direct consequence of the coupling between flavor physics and CP
violation in the KM model.
In this sense, the strong CP problem in KM model is more of a tree
level problem.
All we need is to look for a mechanism beyond Standard model to
suppress the tree level $\theta$ parameter.  In fact, this mechanism
does not have to be a low energy mechanism.  It can be some features
embedded in a high energy theory such as GUT or string theory.  
A popular class of model is the Nelson-Barr
mechanism \cite{Nelson:hg,Barr:qx} in which a (softly broken or gauged)
flavor symmetry and spontaneously broken CP symmetry are used at high
energy in a GUT-like theory to suppress the tree level $\theta$.  The
phase of the KM model is generated by introducing additional heavy
vectorial fermions which can have CP violating mixing with the ordinary
fermions.

In models such as those of Nelson-Barr type, one typically has an
one loop induced $\theta$ at the higher energy scale.  This is of
course because the corresponding tight connection between CP
violation and flavor physics is lost in the extension.  (An
exception can be found in Ref.\cite{bcs} in which this
tight connection is almost maintained). Such contributions are
typically not suppressed by the heavy scale and it is up to the
adjustment of the model parameters to make such contributions
small enough. While flavor symmetry was used in the examples
provided by Ref.\cite{Nelson:hg,Barr:qx}, it can potentially be
replaced by some other symmetry and therefore does not play an
essential role. In addition, in this class of models, the strong
CP $\theta$ parameter receives one loop contribution at high
energy and still needs some fine-tuning to make it small.  Another
recent example\cite{Glashow:2001yz} involving even more direct use
of flavor symmetry, uses the abelian flavor and CP symmetry to
make the up (down) quark mass matrix lower (upper) triangular with
real diagonals.  This will guarantee that tree level $\theta$ is
zero while still have enough parameter to create KM phase of any
magnitude one wishes.  No extra fermions are needed in this type
of models.
%Clearly such mechanism can be easily embedded in GUT or SUSY context.

There are other classes of models that are typified in
Refs.\cite{Masiero:1998yi,Lavoura:1999sk,Babu1990,Chang2003}.  In
particular, Ref.\cite{Masiero:1998yi} uses an $SU(3)$ flavor
symmetry which gives rise to the one-loop suppression of $\theta$.
In Ref.\cite{Lavoura:1999sk}, a flavor sensitive discrete symmetry
is used to reduces the contribution to $\theta$ to the two-loop
level. In Ref.\cite{Babu1990}, left-right gauge symmetry is
employed to reduce the contribution to $\theta$ to the two-loop
level.

In this paper we investigate a few models that can produce KM
model as a low energy effective theory but with naturally small
$\theta$.  The choice of models to review here is of course a
result of our own temporary preference.  We will start with a
review of the models of Glashow, of Masiero-Yanagita with emphasis
on the flavor aspect of the model. Then, we shall present a new
model based on an SO(3) horizontal symmetry.  We will show that in
both $SO(3)$ and $SU(3)$ models, the $\theta$ can be reduced to
two-loop level by using an additional discrete symmetry which can
be broken softly or spontaneously.  A more detailed exposition of this new
class of models is contained in Ref.\cite{Chang2003}.

\section*{Glashow's Model with \\Triangular Quark Mass Matrices}

Glashow invented a softly broken $U(1)$ flavor symmetry to produce a
triangular mass
matrix for quarks at the tree level using three Higgs doublets and soft CP
breaking.
The three Higgs doublets, $H_{(0)}$, $H_{(1)}$ and $H_{(2)}$,
couple to the quarks in a way preserving an abelian
flavor quantum number $F$ given by
$$ F(H_{(k)})=k \ ,\qquad
F(u^{(i)}_R)=F(d^{(i)}_R)=F(q^{(i)}_L)=f(i)\ ,\quad $$
$$
f(i)\equiv\left\{
\begin{array}{lr}
+1 \ ,& \hbox{for } i=1 \\
\ 0\ ,& \qquad      i=2 \\
-1 \ ,&             i=3 \end{array}\right.   \ . $$
Here the quark field is indexed by its generation.

The symmetry turns the Yukawa Hamiltonian into the form:
$$
 y_d^{(i,j)} \overline q_L^{(i)} H_{(j-i)} d_R^{(j)}
+y_u^{(i,j)} \overline q_L^{(i)}(i\tau_2)( H_{(i-j)} )^*  u_R^{(j)}
$$
There are no Higgs field with $F = -1$ or $F = -2$, so the Yukawa
couplings are of the form:
$$
 ( \bar d_L \  \bar s_L \  \bar b_L \ )
\left(\begin{array}{ccc}
y_d^{(1,1)} H_0 &   y_d^{(1,2)} H_1  & y_d^{(1,3)} H_2 \\
         0      &   y_d^{(2,2)} H_0  & y_d^{(2,3)} H_1 \\
         0      &    0               & y_d^{(3,3)} H_0 \end{array}\right)
\left(\begin{array}{c}  d_R \\ s_R \\ b_R \end{array}\right)
$$
%%%
$$ \   +  ( \bar u_L \  \bar c_L \  \bar t_L \ )
\left(\begin{array}{ccc}
y_u^{(1,1)} H_0^* &   0   & 0 \\
y_u^{(1,2)} H_1^* &   y_u^{(2,2)} H_0^* & 0 \\
y_u^{(1,3)} H_2^* &   y_u^{(2,3)} H_1^* & y_u^{(3,3)} H_0^*
\end{array}\right)
\left(\begin{array}{c}  u_R \\ c_R \\ t_R \end{array}\right) \ .
$$
The CP symmetry is assumed to be only softly broken such that the Yukawa
couplings as well as the tree level $\theta$ coupling are both zero.
The vev's of $H$'s produce the mass matrices for the down-type quarks and
the up-type quarks in the form of

$$    (\bar d_L^{(i)}) (M_D)_{ij} (d_R)^{(j)} \ ,\qquad
      (\bar u_L^{(i)}) (M_U)_{ij} (u_R)^{(j)} \ ,   $$
$$ (M_D)=
\left(\begin{array}{ccc}
y_d^{(1,1)} \vev{H_0} &   y_d^{(1,2)} \vev{H_1}  & y_d^{(1,3)} \vev{H_2} \\
         0            &   y_d^{(2,2)} \vev{H_0}  & y_d^{(2,3)} \vev{H_1} \\
         0            &    0                     & y_d^{(3,3)} \vev{H_0}
\end{array}\right) $$
$$ \quad
\equiv \left(\begin{array}{ccc}
m^{(0)}_d            & \epsilon_{12} & \epsilon_{13} \\
0              &  m^{(0)}_s          & \epsilon_{23} \\
0              &  0            & m^{(0)}_b           \end{array}\right) \ ,
$$
%%%
$$ (M^{(0)}_U)=
\left(\begin{array}{ccc}
y_u^{(1,1)} \vev{H_0^*} &   0   & 0 \\
y_u^{(1,2)} \vev{H_1^*} &   y_u^{(2,2)} \vev{H_0^*} & 0 \\
y_u^{(1,3)} \vev{H_2^*} &   y_u^{(2,3)} \vev{H_1^*} & y_u^{(3,3)} \vev{H_0^*}
\end{array}\right)  $$
%%%%%%%%%
$$ \
\equiv \left(\begin{array}{ccc}
m^{(0)}_u            &  0               & 0 \\
\epsilon_{21}^*&  m^{(0)}_c             & 0 \\
\epsilon_{31}^*&  \epsilon_{32}^* & m^{(0)}_t           \end{array}\right) \ ,
$$
where $m^{(0)}_i$ are the zeroth order quark mass without the mixing
effects.  Among the three vev's of the Higgs bosons, $\vev{H_0}$ is chosen
to be real by convention.
$\epsilon_{12}$, $\epsilon_{23}$,
$\epsilon_{21}$ and $\epsilon_{32}$ all
have the same phase and,
$\epsilon_{13}$ and $\epsilon_{31}$ have another.  Since
both flavor and CP are assumed to be
softly broken, one can use the flavor symmetry to remove another phase
from the vev's.  For example, one can make the phase of $\epsilon_{12}$
and
$\epsilon_{23}$ vanish and be left with complex phase only in
$\epsilon_{13}$ and $\epsilon_{31}$.  Note that one special case is that
the phase of $\epsilon_{13}$ is twice that of $\epsilon_{23}$.  In that
case, using the flavor symmetry one can make all vev's real and CP
violation disappears from the mass matrices (even though there are still
soft CP violation in the Higgs mixing).

In the above choice of the $F$ quantum number, we made
$M_U$ lower triangular and $M_D$ upper triangular.
However, if we make another choice,
$$F(u^{(i)}_R)=F(d^{(i)}_R)=F(q^{(i)}_L)=-f(i)\ ,$$
we can have $M_U$ upper triangular but $M_D$ lower.  Both are
phenomenologically feasible but we take the first choice for its
simplicity.  With this choice, one then diagonalizes the mass matrix
perturbatively and derives the mixing angles
$$V_{12}={\epsilon_{12}m_s\over m_s^2-m_d^2}
       - {\epsilon_{21}^*m_u\over m_c^2-m_u^2}
\approx {\epsilon_{12}\over m_s}
       - {\epsilon_{21}^*m_u\over m_c^2} $$

$$V_{23}={\epsilon_{23}m_b\over m_b^2-m_s^2}
       - {\epsilon_{32}^*m_c\over m_c^2-m_t^2} \approx
         {\epsilon_{23}\over m_b}
       - {\epsilon_{32}^*m_c\over m_t^2}
$$

$$V_{13}={\epsilon_{13}m_b\over m_b^2-m_d^2}
       - {\epsilon_{31}^*m_u\over m_t^2-m_u^2} \approx
         {\epsilon_{13}\over m_b}
       - {\epsilon_{31}^*m_u\over m_t^2}
$$
It is clear that the down-quark mass matrix provides the dominant
contribution to the mixing angles, with
$\epsilon_{12} \approx 25 MeV$,
$\epsilon_{13} \approx 13 MeV$ and
$\epsilon_{23} \approx 150 MeV$.
We can choose the convention that
$\epsilon_{12}$, $\epsilon_{23}$ are real,
and only $\epsilon_{13}$ complex in the way consistent to the
Wolfenstein (Chau--Keung) parameterization.  In this convention, the
complex phase in the $(\rho,\eta)$ plane is the phase of $\epsilon_{13}$,
which can be of order ONE!

One should also note that the flavor Abelian group can be reduced from a
continuous phase $e^{iF\phi}$ to discrete phase angles,
$e^{i{2\over6}\pi F}$, while achieving the same kind of triangular mass
matrices.

The above triangular structures are of course not preserved at the one
loop level, and as a result, there will be one loop correction which was
estimated\cite{Glashow:2001yz} to be $\Delta \theta = (1/4\pi)^2
(\epsilon_{13}\epsilon^*_{23}\epsilon^*_{21}/v_2^2 M_u)K$, where $K$ is an
one loop integral factor of order one or smaller.  This can be about
$10^{-9}$ or smaller.  Now, if one takes the masses of the exotic (beyond
SM) scalar bosons, as well as the scale of the dimensionful soft flavor
breaking
scalar couplings, to be around a high energy scale $\Lambda$, the
resulting one-loop $\theta$ depends on $\Lambda$ only logarithmically
through $K$.
Therefore, while these exotic scalar bosons can be as light as $TeV$, it
is not necessary, as far as solving strong CP problem is concern.  One can
imagine taking $\Lambda$ to be a very high scale, these exotic scalar
bosons all decouple and one is left effectively, at low energy, with a
Standard KM Model with small tree level $\theta$.

It is interesting to note that while the model  does not explain the fermion
hierarchy problem, it does tie up the fermion hierarchy problem with the
strong CP problem in the sense that, the smallest of one-loop $\theta$ is
related to the smallness of off-diagonal $\epsilon$ values.

\section*{Hermitian Mass matrix}

In the model proposed by Masiero and Yanagida\cite{Masiero:1998yi},
the existing three generations are extended
with another three generations. The new three generations are
individually $SU_L(2)$
singlets and vector-like; nonetheless, the hypercharges are chosen
such that they have same electric charges as the known
generations.  Therefore they are labelled by
$$U_{Li}, U_{Ri}, D_{Li}, D_{Ri} \ ,$$
in an analogous fashion   with the known quarks,
$$ q_{Li} \equiv \left(\begin{array}{c} u_{Li} \\  d_{Li}\end{array}\right),
      u_{Ri},d_{Ri} \ .$$
The only difference between the two sets of fermions is in $SU_L(2)$ assignments.
A horizontal flavor symmetry $SU(3)_\parallel$ transforms every Weyl
fermion multiplet above in the fundamental representation,
labelled by generation index $i=1,2,3$.
There are new neutral (and inert to $SU(2)_L \times U(1)$) Higgs
bosons: three $SU(3)_\parallel$ octets $\phi^a_\alpha$ ($a=1,\cdots,8$, $\alpha =1,2,3$) and
a singlet $\Phi$.  The Yukawa couplings in the Hamiltonian is
\begin{eqnarray*}
\bar d_R(g_{d\alpha}\phi^a_\alpha \lambda^a+g'_d\Phi) D_L
                   \\
+\bar
u_R(g_{u\alpha}\phi^a_\alpha \lambda^a+g'_u\Phi) U_L
\\
+ h_d \bar D_R  H^\dagger q_L
+ h_u \bar U_R  \tilde H^\dagger  q_L + {\rm H.c.}
\end{eqnarray*}
with the usual $SU_L(2)$ Higgs doublet $H$ which couples to fermions in
a way crossing two kinds of fermions.  We denote
$ \tilde H \equiv i\tau_2 H^* $.

An alert reader will tell that a criterion  is need to distinguish
$u_R$ from $U_R$, or $d_R$ from $D_R$.  Ref.\cite{Masiero:1998yi} imposed a discrete symmetry
under which $u_R, d_R$ and all $\phi$, $\Phi$ fields are odd, while $U$
and $D$ are even.
The vev's of $\phi$'s and $\Phi$ and $H$ give rise to the
$6\times 6$ mass matrix term
$$ (\bar d_R \  \bar D_R )
\left(\begin{array}{cc}    0     &   g_d\vev\phi^a\lambda^a+g'_d\vev\Phi\\
                         h_d\vev{H^\dagger} & M_D  \end{array}\right)
\left(\begin{array}{c} d_L \\ D_L \end{array}\right) \ . $$
Here $M_D$ is the heavy allowed vector mass of the $D$ field.
The $6\times 6$ matrix has real determinant when couplings are real,
as required by the imposed CP symmetry.
Note that the singlet is necessary, otherwise the Hermitian mass matrix
will be traceless which is no allowed phenomenologically. 
Also note that to break CP and to have most general Hermitian mass matrix, one
need all components of the octets to develop VEV, this is why it is
necessary for have three octets.  It is not clear whether this has been
proved in the literature or not.

Integrating out the exotic generation $D$, we have the effective mass
matrix for the known generations,
$$ (1/M_D) h_d \vev{H^\dagger} (g_d\vev{\phi^a}\lambda^a+g'_d\vev\Phi) \ , $$
which is Hermitian with real determinant.
Some components of Gellmann matrices $\lambda^a$ are complex to produce
the desired CKM phenomenology of CP violation.

The effective mass can be understood as the amplitude given by the
diagram below. It is a kind of see-saw mechanism in the quark sector.
Similar formulas occur for the up-type quarks.
\def\dum{
\begin{center}
\begin{picture}(100,120)(50,0)
\ArrowLine(0,0)(50,50)         \Text(25,25)[lt]{\ $d_L$}
\ArrowLine(50,50)(100,50)      \Text(75,45)[t]{\ $D_R$}
\Line(95,45)(105,55) \Line(95,55)(105,45)  \Text(100,59)[b]{$M_D$}
\ArrowLine(100,50)(150,50)     \Text(125,45)[t]{\ $D_L$}
\ArrowLine(150,50)(200,0)      \Text(175,25)[lt]{\quad $d_R$}
\DashArrowLine(50,50)(0,100){5}\Text(25,75)[lt]{\quad$H$}
\DashArrowLine(200,100)(150,50){5}\Text(175,75)[lt]{\ $\phi, \Phi$}
\Text(150,55)[rb]{$g$}
\Text(55,55)[lb]{$h_d$}
\end{picture}
\end{center}
}
\begin{figure}
\centering
\includegraphics[width=8cm]{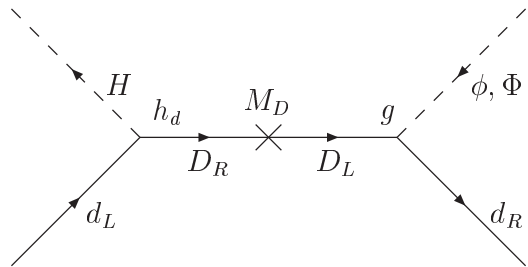}
\caption{\label{fig1} A see-saw diagram for the effective masses of
 the known 3 generation. }
\end{figure}
It is interesting to note that the mass matrix is Hermitian certainly
because the $SU(3)$ symmetry plus spontaneous CP violation, the fact
that $\phi^\alpha$ are octet.  However even after SU(3) is broken, the
hermiticity of the mass matrix is still maintained. 

\section*{strong CP $\theta$ issue}
As long as the mass matrix is block-Hermitian, 
the contribution to $\theta$ will
vanish.  So, to look for contribution to $\theta$, one looks
for loop induced operator that may violate this hermiticity, such as
$$(1/M^n) \bar{d}_L d_R H f_n(\phi, \Phi)  $$
where $f_n$ is a function of $\phi$ or $\Phi$.  The discrete symmetry requires function $f_n$ to be of even power.  However, since the fermion
bilinear can only either be $SU(3)$ singlet or octet, $f_n$ has to be
effective singlet or octet.  Naively it may seem that there will be a one loop contribution to $\theta$ at this point.  However, as shown in Ref.\cite{Chang2003}, the one loop $\theta$ remains zero in this model.  

Note that Higgs scalar potential cannot break parity.  It can be done only by chiral fermions.
But since P is already broken by the chiral fermion in SM, violation of
charge conjugation in the Higgs VEV produces CP violation.
Ref.\cite{Masiero:1998yi} used the nonabelian character of SU(3) to break charge conjugation, $C$, in the Higgs potential.  In principle this can be done by any irrep
in nonabelian group as long as one uses enough copies.  We call this "nonabelian CP violation".  
One may wonder what is the simpler nonabelian model that can achieve the same goal.  The model in Ref.\cite{Chang2003} may be the simplest example one can find.  

\section*{Minimal model of this Class}

In Ref.\cite{Chang2003}, the flavor symmetry is replaced by smaller $SO(3)$ symmetry.  The new three generations are
individually $SU_L(2)$
singlets and vector-like, nonetheless, the hypercharges are chosen
that they have same electric charges as the known
generations. Therefore they are labelled by
$$U_{Li}, U_{Ri}, D_{Li}, D_{Ri} \ ,$$
in an analogous fashion   with the known quarks,
$$ q_{Li} \equiv \left(\begin{array}{c} u_{Li} \\  d_{Li}\end{array}\right),
      u_{Ri},d_{Ri} \ .$$
A horizontal flavor symmetry $SO(3)_\parallel$ transforms every Weyl
fermion multiplet above in the ${\BM 3}$ representation,
labelled by generation index $i=1,2,3$.
There are new horizontal neutral (inert to $SU(2)_L \times U(1)$) Higgs
bosons, {\it i.e.} one quintet (symmetric traceless rank-2 tensor)
CP-even  $\phi_S$
and one triplet (antisymmetrc rank-2 tensor) CP-odd $\phi_A$.
The Yukawa couplings are
\begin{eqnarray*}
   \bar  d_R(\mu_d+g_{dS}\phi_S +ig_{dA}\phi_A) D_L \\
 + \bar  u_R(\mu_u+g_{uS}\phi_S +ig_{uA}\phi_A) U_L
\\
 + \bar  D_R(\mu_D+g_{DS}\phi_S +ig_{DA}\phi_A) D_L  \\
 + \bar  U_R(\mu_U+g_{US}\phi_S +ig_{UA}\phi_A) U_L  \\
+ (h_d \bar d_R+h'_d \bar d_R) H^\dagger q_L \\
+ (h_u \bar u_R+h'_u \bar U_R) \tilde H^\dagger  q_L + {\rm H.c.}
\end{eqnarray*}
with the usual $SU_L(2)$ Higgs doublet $H$ which couples to fermions
flavor-blindly.
We denote  $ \tilde H \equiv i\tau_2 H^* $.

The vev's of $\phi_S$ and $\phi_A$ and $H$ give rise to the following
$6\times 6$ mass matrix term
$$ (\bar d_R \  \bar D_R ) (M_6)
\left(\begin{array}{c} d_L \\ D_L \end{array}\right) \ , $$
%%%%%%%%%
$$ (M_6)=
\left(\begin{array}{cc}
  h_d\vev{H^\dagger}\BM{1} & \mu_d + g_{dS}\vev{\phi_S} +ig_{dA}\vev{\phi_A}\\
 h'_d\vev{H^\dagger}\BM{1} & \mu_D + g_{DS}\vev{\phi_S} +ig_{DA}\vev{\phi_A}
  \end{array}\right)
$$
The $6\times 6$ matrix has real determinant when couplings are real,
as required by the imposed CP symmetry.
Contrary to the $SU(3)$ model, it is not necessary to have a $SO(3)$ singlet
if one does not impose a discrete symmetry.

Integrating out the exotic generation $D$, we have the effective mass
matrix for the known generations like CKM phenomenology.
The effective mass can be understood as the amplitude given by the
diagram below. It is a kind of see-saw mechanism in the quark sector.
Similar formulas occur for
the up-type quarks.
\def\dum2{
\begin{center}
\begin{picture}(100,120)(50,0)
\ArrowLine(0,0)(50,50)         \Text(25,25)[lt]{\ $d_L$}
\ArrowLine(50,50)(100,50)      \Text(75,45)[t]{\ $D_R$}
\Line(95,45)(105,55) \Line(95,55)(105,45)
\ArrowLine(100,50)(150,50)     \Text(125,45)[t]{\ $D_L$}
\ArrowLine(150,50)(200,0)      \Text(175,25)[lt]{\quad $d_R$}
\DashArrowLine(50,50)(0,100){5}\Text(25,75)[lt]{\quad$H$}
\DashLine(200,100)(150,50){5}\Text(175,75)[lt]{\ $\phi_S, \phi_A$}
\Text(150,55)[rb]{$g_{dS,A}$}
\Text(55,55)[lb]{$h'_d$}
\end{picture}
\end{center}
}
\begin{figure}
\centering
\includegraphics[width=8cm]{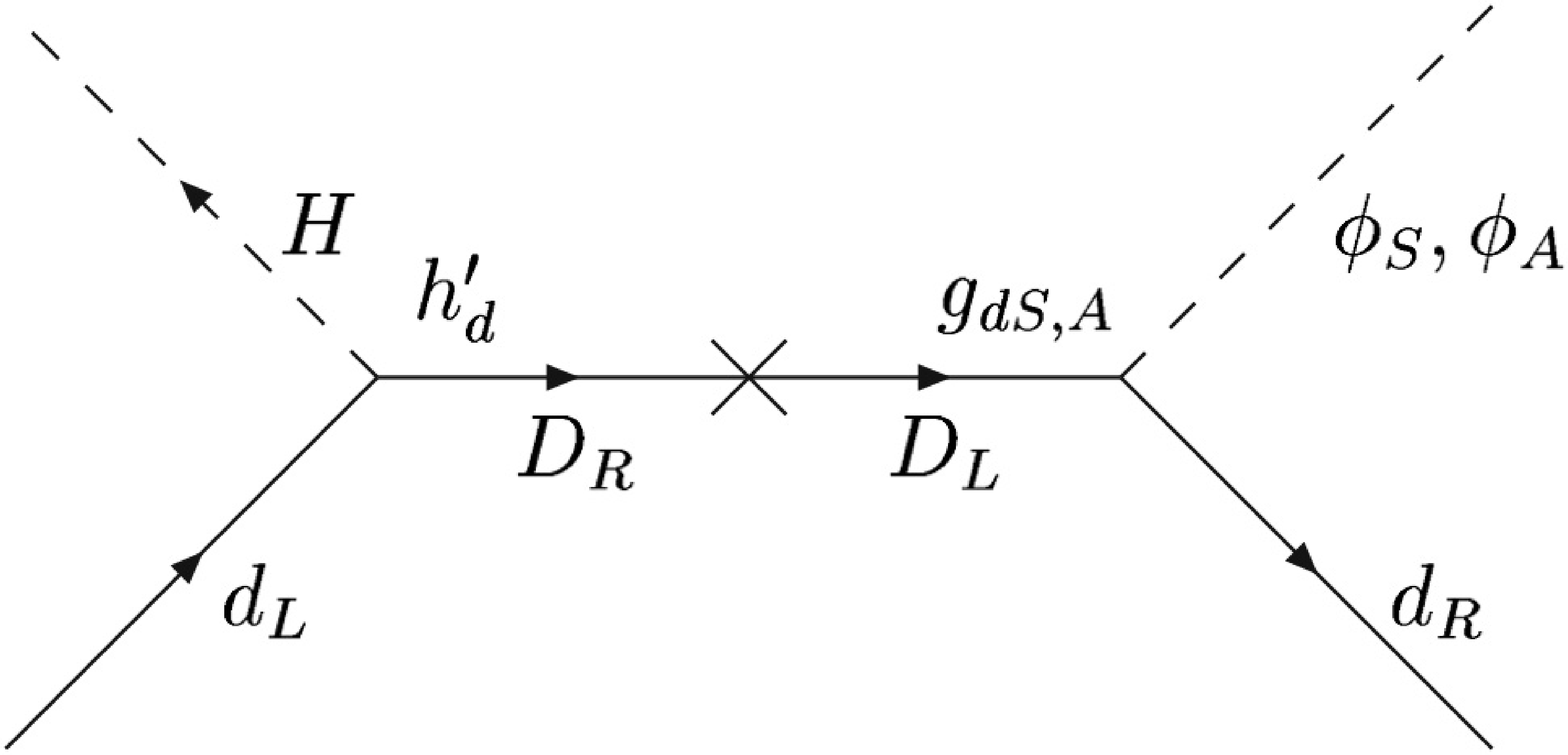}
\caption{\label{fig2} A see-saw diagram for the effective masses of
 the known 3 generation. }
\end{figure}
The low energy effective mass matrix can be obtained through the
following steps.  In the limit of $\vev{H}=0$, $d_L$ quarks decouple
and we have the reduced mass matrix of a size $6\times3$ instead,
$$ (\bar d_R \  \bar D_R )
\left(\begin{array}{c} M_d\\M_D
  \end{array}\right)
\left( D_L \right)
\ ,$$
%%%%%%%%%%%%
$$ \qquad \left.\begin{array}{l}
    M_d=  \mu_d +   g_{dS}\vev{\phi_S} +ig_{dA}\vev{\phi_A} \ ,\\
    M_D=  \mu_D +   g_{DS}\vev{\phi_S} +ig_{DA}\vev{\phi_A} \ .
\end{array}\right.
$$
We find a $6\times 6$ unitary matrix $V$ to transform the the above
$6\times3$ into one with zero entries in the upper $3\times 3$ block.
$$ V \left(\begin{array}{c} M_d\\M_D\end{array}\right)=
     \left(\begin{array}{c} \BM{0}  \\M'_D\end{array}\right)  \ , $$
by choosing the top three row vectors of $V$
perpendicular to the 3 column vectors in the mass matrix. It is
possible because the 6 dimensional linear space is
larger than the 3 dimensional space spanned by the three column
vectors.
Furthermore, by using bi-unitary transformation,
we also rotate $D_L$  into $D'_L$
so that  $M'_D$ is diagonal. In this way, the massless states $d'_R$ and
the massive states $D'_R$ are generally  mixed  among the original $d_R$ and
$D_R$. Nonetheless, the three generations of $d_L$ remain massless and unmixed.

We include the effect $\vev{H}$ from the viewpoint of
perturbation. The mass terms involving $d_L$ are tabulated in the
matrix form,
$$
(\bar d'_R \  \bar D'_R ) V
\left(\begin{array}{c}
  h_d\vev{H^\dagger}\BM{1}   \\
 h'_d\vev{H^\dagger}\BM{1}  \end{array}\right)
\left( d_L \right)
=(\bar d'_R \  \bar D'_R )
\left(\begin{array}{c}
  {\hat{m}}_d \\
  {\hat{m}'_d}
\end{array}\right) \left( d_L \right)   \ .
$$
Including  $D'_L$, we have
$$
(\bar d'_R \  \bar D'_R )
\left(\begin{array}{cc}
  {\hat{m}}_d  & \BM{0}\\
  {\hat{m}'_d} &  M'_D
\end{array}\right)
\left( \begin{array}{c} d_L\\ D'_L\end{array} \right)   \ .
$$
As $M'_D\gg\hat{m}_d$,
masses of  usual $d$-quarks are basically given by
diagonalization of the mass matrix $\hat{m}_d$.
Similar procedures also apply to the $u$-quarks.
Phenomenology of CKM mechanism follows.

The model as it is will have one-loop contribution to $\theta$.  To
make the one-loop contribution vanishes, it is necessary to make the
$\bar{d_R}d_L$ block in the mass matrix $M_6$ vanishes at tree level.
This can be easily done with a discrete symmetry as explained in
Ref.\cite{Chang2003}.  This will make the $\theta$ of the low energy
effective KM model naturally small.  Note that the smallness is not
due to the suppressive effect of any high energy scale.  It is a
consequence of natural smallness of the two loop quantum effect.

\section*{Comparison with Nelson-Barr scheme}

Let us make a brief comparison with
Nelson-Barr\cite{Nelson:hg,Barr:qx} scheme before we conclude.  They
proposed the model in grand unified theory (GUT) context.  To compare,
we first  strip the model off the GUT context.  For the extra heavy
fermions, they use an additional set of Standard Model fermions plus
its parity mirror (to make it a vectorial set).  They impose flavor
symmetry on the SM fermions, but not the extra fermions.  Therefore
their heavy fermions are flavor singlets and one needs to use only one
copy, that is, 
$$Q_{L}\equiv \left(\begin{array}{c} U_{Li} \\  
D_{L}\end{array}\right), U_{R}, D_{R} , 
Q'_{R}\equiv \left(\begin{array}{c} U'_{R} \\
D'_{R}\end{array}\right), 
U'_{L}, D'_{L} \ ,$$
in addition to the known quarks,
$$ q_{Li} \equiv \left(\begin{array}{c} u_{Li} \\  d_{Li}\end{array}\right),
      u_{Ri},d_{Ri} \ .$$
The flavor symmetry is broken by a set of $SU(2)$-singlet Higgs fields
$\Phi,\Phi'$ that connect the light and heavy fermions. 
The vev's of $\Phi$ and $H$ give rise to the
$5\times 5$ mass matrix term
$$ (\bar d_{Ri} \  \bar D'_R \ \bar D_R)
\left(\begin{array}{ccc}    \lambda\vev{H^\dagger}     & 0 &   f \vev\Phi\\
  f' \vev{\Phi'} & M_{D_1}     & 0 \\
      0 & 0  & M_{D_2}  \end{array}\right)
\left(\begin{array}{c} d_{Li} \\ D_L \\ D'_L \end{array}\right) \ . $$
Here $M_{D_{1,2}}$ are the heavy Dirac masses of the $D$ and $D'$
fields. 
CP symmetry is arranged such that it is broken in the off-diagonal  
$f \vev{\Phi}$ and 
$f' \vev{\Phi'}$ components.
Note that these components do not contribute to the mass determinant.
The $5\times 5$ matrix has real determinant when couplings are real,
as required by the imposed CP symmetry.  
Clearly, one can see the difference between the Nelson-Barr mechanism
and the block-Hermitain mechanism.

\section*{Conclusion}

There are many solutions to the strong CP problem.  Axionic models are
among the most popular.  It is not difficult to include axion in grand
unified theories or even string theory.  However, it may be
interesting to think that strong CP problem may be our best hint to
some unknown high energy physics.  One of the other current puzzles of
particle physics which we also wish to be resolved by some unknown
high energy physics is the flavor problem including the mass hierarchy
problem.  One may wonder if the two puzzles may have the same solution
which is the flavor symmetry.  In this paper we explore this
possibility to some extend.  Certainly there are a lot more
possibilities to cover.

\vskip 0.1in
\noindent
{\it Acknowledgment.} WYK  is supported by the
U.S.~DOE under Grants No.~DE-FG02-84ER40173.  DC is supported by a
grant from NSC of Taiwan, ROC.  DC also wishes to thank the COSPA
Center at NTU, Theory Group at SLAC and Theory Group at LBL for
hospitality during this work.

\end{document}